# Efficient molecule discrimination in electron microcopy through an optimized orbital angular momentum sorter

F. Troiani[1], E. Rotunno[1,*], S. Frabboni[2], R.B.G. Ravelli[3], P.J. Peters[3], E Karimi[4], V. Grillo[1]

[1] Centro S3, CNR-Istituto di Nanoscienze, via G. Campi 213/A, I-41125 Modena, Italy
[2] University of Modena and Reggio Emilia, via G. Campi 213/A, I-41125 Modena, Italy
[3] The Institute of Nanoscopy, Maastricht University, 6211 LK Maastricht, The Netherlands
[4] Department of Physics, University of Ottawa, Ottawa, Ontario, K1N 6N5, Canada

**Abstract.** We consider the problem of discriminating macromolecular structures in an electron microscope, through a specific beam shaping technique. Our approach is based on maximizing the which-molecule information extracted from the state of each electron. To this aim, the optimal observables are derived within the framework of quantum state discrimination, which allows one to fully account from the quantum character of the probe. We simulate the implementation of such optimal observable on a generalized orbital angular momentum (OAM) sorter, and benchmark its performance against the best known real-space approach.

The quest for extreme sensitivities in the investigation of atomic and molecular systems naturally leads to the use of single particles as quantum probes. The state of such probes is expected to display a strong dependence on the system under investigation, and thus to allow an efficient inference on the system of interest. However, the use of single-particle probes also implies inherent limitations, resulting from fundamental quantum features, such as decoherence and the measurement back action. Decoherence tends to smear out the information encoded in the particle state, while the measurement back-action limits the amount of information on the state of a quantum system that can actually be extracted, and results in the impossibility of discriminating two non-orthogonal states. A full account of these quantum aspects is thus required in order to identify the optimal measurement strategy.

In electron microscopy the quantum probe is represented by the electron. The measurement process implies the "collapse" of its wave function, which reduces to a point (pixel) on a detector. A full image of the sample is obtained by using a large number of electrons, whose distribution on the detector represents the measurement statistics[1,2]. However, the number of electrons used in the imaging of systems such as proteins should be restricted, in order to limit radiation damage. This has fueled the development of novel techniques, such as cryo-microscopy and single particle analysis[3]. On the other hand, these techniques require the imaging of many supposedly identical proteins, while the accurate imaging of a single protein remains largely prohibitive. The recent introduction of beam shaping techniques[4,5] has changed the way in which electron microscopy can be afforded, and the concept of imaging itself. In particular, through the use of appropriate electrostatic elements[6,7,8] or holograms[9], it is possible to analyze the wave function in different bases, and thus to implement the measurement of different observables. The most considerable and already demonstrated case is that of the orbital angular momentum (OAM) sorter[10], which is inspired by optics[11]. The large flexibility that this approach allows in the choice of the measured observable represents a key resource for maximizing, in a system-specific fashion, the amount of information that can be extracted from the electron wave function, thus reducing the number of probes that are required to achieve a given degree of confidence in the final inference.

In the present manuscript, we consider the problem of identifying a protein by means of an OAM sorter. In order to fully exploit the potentialities of this approach, on the one hand, we refine the sorting by including additional electro-optical elements, which implement a correlated measurement of the radial and angular degrees of freedom. On the other hand, we fully account for the quantum nature of the probe (electron), by investigating the problem within the framework of quantum state discrimination. This allows us to derive the optimal measurement strategy, both in the ideal case of a full knowledge on the alternative protein states and coherent electron dynamics, and in the more general case where the state of the molecule is partially unknown and/or the electron dynamics is affected by decoherence. In particular, we show that the OAM sorter with an additional projection element implements the optimal measurement strategy in the representative cases of decoherence in the angular degrees of freedom or of a complete lack of knowledge on the protein orientation.

The prototypical problem we consider can be schematically summarized as follows. There are two hypotheses, hereafter labeled "0" and "1", concerning different features of the physical system of interest. In particular, we consider the case where the system is a protein, each hypothesis $I_k$ (with $k=0,1$) specifies its identity $X_k$ and orientation, and is assigned an a priori probability $p_k$ (for example given by the dilution of the two proteins or by the likelihood of a model). Each electron that is used as a quantum probe interacts with the molecule, and is left either in the state $|\psi_0\rangle$ or in $|\psi_1\rangle$, depending on which of the two hypotheses applies. A measurement is eventually performed in order to identify the electron state, and thus to infer the correct $I_k$.

The above problem can be formalized within the general framework of quantum state discrimination[12,13]. For the sake of the following discussion, it's convenient to expand the two alternative electron states in terms of the normalized projections $|m, \chi_{k,m}\rangle$ on the eigenspaces of the angular momentum $L_z$ ($z$ is the electron-propagation direction):

$$|\psi_k\rangle = \sum_m \sqrt{q_{k,m}} e^{i\alpha_{k,m}} |m, \chi_{k,m}\rangle, \quad (1)$$

where $0 \leq q_m \leq 1$ is the probability associated to each value $m$ of $L_z$ and $\chi_{k,m}$ represents the radial component of the state.

The association of a pure electron state with each of the two hypotheses is based on the assumption that the electron dynamics is coherent and that the protein state is perfectly defined within each $I_k$. If instead the state of the protein is partially unknown and/or the electron is affected by decoherence (e.g., due to inelastic scattering), the state vectors $|\psi_k\rangle$ have to be replaced by density operators $\rho_k$. In particular, if the molecule orientation around the $z$ axis is completely undefined, as is usually the case, or if the electron decoherence destroys the phase coherence between different eigenstates of $L_z$, the density operators take the form[14]

$$\rho_k = \sum_m q_{k,m} |m, \chi_{k,m}\rangle\langle m, \chi_{k,m}|, \quad (2)$$

where all the information that was contained in the phases $\alpha_{k,m}$ has been erased.

The discrimination between the two electron states is performed on the basis of the measurement outcome. In particular, one identifies two outcomes whose occurrence makes more likely either one hypothesis or the other. Formally, each outcome is associated to a probability (*i.e.* nonnegative and Hermitian) operator $\Pi_k$, whose expectation value gives the probability that the $k$-th outcome occurs. The probability of identifying the correct hypothesis on the basis of a single-electron measurement is thus given by[13,14]

$$p = p_0 \text{tr}(\rho_0 \Pi_0) + p_1 \text{tr}(\rho_1 \Pi_1), \quad (3)$$

where the first (second) term represents the probability that the first (second) hypothesis is true and that the measurement provides the corresponding outcome $k=0$ ($k=1$). The equation also applies to the case of pure electron states, with $\rho_k = |\psi_k\rangle\langle\psi_k|$. The probability $p$ is the figure of merit we refer to in the following, and can be maximized by a suitable choice of the measurement. There is, however, a fundamental limitation related to the quantum nature of the probe: if the two electron states and are not orthogonal, as is generally the case, there is no quantum measurement that can perfectly discriminate them. In fact, in the case of two pure states, the probability $p$ cannot exceed the Helstrom bound<sup>Errore. Il segnalibro non è definito.,Errore. Il segnalibro non è definito.</sup>

$$p_{max}^{\psi} = \frac{1}{2} + \frac{1}{2}[1 - 4p_0p_1|\langle\psi_0|\psi_1\rangle|^2]^{1/2}, \quad (4)$$

which ranges from ½ to 1 as the overlap between the two states varies from 1 to 0. In passing from the $|\psi_k\rangle$ to their dephased counterparts $\rho_k$, the two electron states tend to become less distinguishable and the probability $p$ of discriminating between the two hypotheses thus tends to decrease. The upper bound is now given by[14]:

$$p_{max}^{\rho} = \frac{1}{2} + \frac{1}{2}\sum_{m}\left[(p_0 q_{0,m} + p_1 q_{1,m})^2 - 4p_0 p_1 q_{0,m} q_{1,m}|\langle\chi_{0,m}|\chi_{1,m}\rangle|^2\right]^{1/2}, \quad (5)$$

which reduces to the Helstrom bound if both the $|\psi_k\rangle$ belong to the same eigenspace of $L_z$. The above limits can be achieved by identifying and implementing the quantum measurement that is optimal, given the hypotheses $I_k$ (and thus the electron states $\rho_k$) and the a priori probabilities $p_k$.

For a given measurement, the probability $p$ of identifying the correct hypothesis can be maximized by assigning each outcome to one of the two hypotheses, according to a maximum likelihood criterion. This also applies to the case of measurements with more than two possible outcomes. In particular, we consider the case

$$p_{OAM}^{\rho} = p_{OAM}^{\psi} = \sum_{m}\max\{p_0 q_{0,m}\langle\chi_{0,m}|\pi_{0,m}|\chi_{0,m}\rangle, p_1 q_{1,m}\langle\chi_{1,m}|\pi_{1,m}|\chi_{1,m}\rangle\} \equiv p_{OAM} \quad (6)$$

where $\pi_{k,m}$ are the probability operators corresponding to the two possible outcomes of a radial observable, within each eigenspace of $L_z$. Given the dependence on $m$ of the $\pi_{k,m}$, this corresponds to a correlated measurement of the radial and angular degrees of freedom. As pointed out in the first equality, the discrimination probability is unaffected by the loss of phase coherence between the different angular momentum components $|m, \chi_{k,m}\rangle$. The above measurement strategy can always be made optimal ($p_{OAM} = p_{max}^{\rho}$) by identifying $\pi_{0,m}$ and $\pi_{1,m}$ respectively with the projector on the positive or on the negative eigenstate of $\sigma_m \equiv \sum_{j=0,1}(-1)^j p_j q_{j,m}|\chi_{j,m}\rangle\langle\chi_{j,m}|$. The optimal projectors increasingly differ from those on $|\psi_0\rangle$ and $|\psi_1\rangle$, for increasing overlap between the two electron states. In the following, we shall discuss the implementation of the optimal approach by means of the OAM sorter.

So far we have considered the limiting case where the discrimination between the hypotheses $I_0$ and $I_1$ is performed on the basis of a single-electron measurement. The probability of identifying the correct hypothesis can be increased by repeating such measurement $N$ times, one on each of the electrons that has interacted with the molecule. In this case, the possible outcomes of the overall measurement can be identified with the different number of times $n_0$ in which the outcome 0 occurs. These outcomes follow a binomial distribution, with a success probabilities $s_0$ or $(1 - s_1)$, depending on whether $I_0$ and $I_1$ applies, where $s_k \equiv \text{tr}(\rho_k \Pi_k)$. The expression of the probability for the $N$-electron case is thus given by[14]:

$$P(N) = \sum_{n_0=0}^{N}\binom{N}{n_0}\max\{p_0 s_0^{n_0}(1-s_0)^{N-n_0}, p_1 s_1^{n_1}(1-s_1)^{N-n_1}\}. \quad (7)$$

From the above expression one can derive the minimum number of electrons $N_{min}(x)$ that are required in order to exceed a given threshold $x$ for the probability $P(N)$.

The discrimination strategy that we consider in the following is based on the use of the OAM sorter with an additional phase plate element (Fig. 1). Here, each electron of the beam is prepared in a plane-wave state, before interacting with the protein under investigation. Such interaction perturbs the electron state, in a way that depends on the protein identity and state. Suitably engineered phase elements then direct states corresponding to different values of $L_z$ to different regions of the detector, thus implementing a measurement of the orbital angular momentum. As an important refinement of the measurement strategy, additional phase elements are introduced, which further sort each angular momentum component on the basis of the radial state. The generalized OAM sorter thus implements the correlated measurement of the angular and radial degrees of freedom we refer to above (see Eq. (6) and related discussion).

The electron wave function, and in particular its transverse $(x,y)$ component $\psi_k$, is affected by the interaction with the protein. After such interaction, $\psi_k$ can be written as[15]:

$$\psi_k(x,y) = A(x,y)\exp[i\sigma V(x,y)]\,\psi_{probe}(x,y). \quad (8)$$

Here, $\psi_{probe}$ is the electron wave function before the interaction with the protein $X_k$, which in the plane-wave case is given by a constant, and $\sigma = 2\pi m_0 \gamma \lambda / h^2$ (being $m_0$ the electron rest mass, $\gamma$ the relativistic factor, $\lambda$ the wavelength, and h the Plank constant). The protein modulates the phase and, to a minor extent, the amplitude of the electron wave function. Since the proteins are made of light element, the phase modulation is typically small ($\sigma V \ll \pi$). The overlap between the electron states $|\psi_0\rangle$ and $|\psi_1\rangle$, corresponding to the presence of the two alternative proteins $X_0$ and $X_1$, is thus quite large and, as a result, the theoretical maximum for the discrimination probability given by the Helstrom bound is close to 1/2 .

We apply the above approach to three test protein models. Two of these, hereafter labelled Pa and Pb, refer to the EspB protein of *Mycobacterium Tuberculosis* and are characterized by a 7 –fold symmetry. In particular, Pa is based on computational modelling[16] and is more loosely packed compared to Pb, which is derived from experimental data[17]. The third protein model is similar to the Pa but has a 6-fold symmetry. Experimentally, the discrimination between the last protein model and the previous two represents a very important case, since normally these proteins are hardly distinguishable. Besides, it corresponds to opposite qualitatively different discrimination with respect to that between Pa and Pb, because the difference between the protein structures is mainly in the azimuthal – rather than in the radial – direction. The effect on the electron wave function of the interaction with a protein is visualized in Fig. 2 for the case of Pb. In particular, the interaction induces a phase modulation in the *xy* plane, which is displayed in Cartesian (a) and log-polar coordinates (b). With the OAM sorter, the electron wave function is decomposed into different contributions, corresponding to different values of *m* of $L_z$ (c). The cylindrical lens finally diffracts the radial degree of freedom, while leaving the OAM channels separated (d). For a given protein, the phase element is adapted in order to match and conjugate the dependence of the phase on the radial degree of freedom, and this is done independently for each value of *m*. The result is a "phase flattened" wave function[18], that diffracts nearly exactly to a point. In these conditions, the radial diffraction produces a very sparse representation of the electron wave function for a single target protein. (Ideally, this corresponds to setting $\pi_{0,m} = |\chi_{0,m}\rangle\langle\chi_{0,m}|$, i.e. to identifying the probability operators with the projectors on the electron state components.) We note that these images refer to the case of pure electron states. If these are affected by decoherence or by an undefined protein orientation, the angular dependence of the phase modulation displayed in the upper panels is entirely blurred away, whereby the phase modulation at each point is replaced by the corresponding angular average. Instead, the patterns obtained by means of the OAM sorter and displayed in the lower panels are unaffected by dephasing [14].

The simulation of the protein discrimination procedure, based on the use of the optimal observable, is reported in Fig. 3 for the case of Pa and Pb. The simulations refer to the case of limited doses (2 and 0.2 e/Å$^2$) and are performed by double extraction Montecarlo methods. As expected, the optimal observables don't give rise to a complete localized in the radial basis [panels (a,b,d,e)], because in this case the probability operators $\pi_{k,m}$ project on a basis that is diagonal with respect to that formed by $|\chi_{0,m}\rangle$ and $|\chi_{1,m}\rangle$. This is particularly true for the $m$=0 subspace (c,f), where the largest component of the electron wave functions is concentrated. In the $m\neq0$ subspaces, the radial components are often nearly orthogonal (some are even forbidden by the different symmetries), and the optimal observables are approximately given by $\pi_{k,m} = |\chi_{k,m}\rangle\langle\chi_{k,m}|$. As a general comment, we note that our method produces a sparse distribution of the detected electrons (i.e. most population is concentrated in a few pixels) and is therefore less affected by noise as compared to approaches where this is not the case.

In order to benchmark and quantitatively compare the sorter-based discrimination strategies, we compute the relevant probabilities (Table 1). We start by quantifying the which-protein information encoded in the electron state (columns 2-4). As anticipated above, the overlap $|\langle\psi_0|\psi_1\rangle|$ between the electron wave functions corresponding to the two hypotheses is close to 1 for all pairs of proteins. Correspondingly, a single-electron probe allows a maximal discrimination probability ($p_{max}^{\psi}$) that is a few cents above the "blind-guess" value of ½, and is slightly reduced ($p_{max}^{\rho}$) by dephasing.

We then quantify the suitability of the different measurements to access such which-protein information, by means of the corresponding probabilities $p$ (columns 5-6). For the imaging in real space (RS), we also considered an ideal Zernike phase plate[19,20] that introduces a $\pi/2$ phase shift at the center of the diffracted image. We note that such phase shift has never been exactly realized in practice, so that the reported values should be regarded as upper bounds for the imaging in real space approach. Even in such ideal case, the obtained values of the discrimination probabilities for the state ($p_{RS}^{\rho}$) fall below the corresponding theoretical maxima. The optimal observables implemented through the OAM sorter allow for the achievement of discrimination probabilities ($p_{OAM}$) that are comparable and in most cases significantly larger than the those achieved with the ideal Zernike phase plate,. The small difference between $p_{OAM}$ and the theoretical maximum $p_{max}^{\rho}$ (which should in principle be zero) is possibly due to imperfections in the phase-flattening based implementation of the optimal projectors $\pi_{k,m}$.

The maximal suitability of the implemented observable to distinguish between the two alternative electron states results in a minimization of the number of particles that are required in order to achieve a threshold value $x$ for the discrimination probability. The reported values of $N(x)$, all of the order of $10^2$, correspond to doses of much less than 0.1e /Å$^2$ (columns 7-9). This number is quite small and typically at least 50% better the case of the ideal Zernike phase plate. We calculated that when the phase introduced by the protein is increased by a factor 3, for example by introducing thicker proteins or through the recently introduced multi-pass approaches [21], the advantage dose reduction allowed by the optimal observable is of even an order of magnitude.

Moreover within this approach and with a typical affordable dose of few e /Å$^2$, something like 100 tests between different pairs of protein can be carried out simply by changing the phase of the final sorting element in a programmable way[22,23]. This implies a high degree of flexibility in the implementation of the OAM-based approach, and suggests the possibility to use adaptive learning in cases where the possible identities of the protein are not limited to two options, as assumed above.

In conclusion, we have investigated the problem of discriminating two proteins by means of electron microscopy. The discrimination procedure has been investigated within the framework of quantum state discrimination, which allows us to fully account for the quantum nature of the electron. The discrimination probability based on the use of the angular momentum sorter is unaffected by a dephasing process resulting from the electron decoherence and/or from an uncertainty on the protein orientation. We have shown that the generalized OAM sorter, which implements a correlated measurement of the radial and angular degrees of freedom, can in principle realize an optimal discrimination strategy, and have provided a concrete example of such implementation with a benchmark on models of proteins with different radial and/or azimuthal phase distribution. This optimization represents a fundamental means for minimizing the number of probes (electrons) that are required for the protein identification, and thus for limiting the induced damage.

**Acknowledgments**

This work is supported by Q-SORT, a project funded by the European Union's Horizon 2020 Research and Innovation Program under grant agreement No. 766970.- Q-SORT (H2020-FETOPEN-1-2016-2017). We acknowledge Abril Gijsbers, Ye Gao and Axel Siroy (UM) for having provided the EspB model.

| $X_0, X_1$ | $\|\langle\psi_0\|\psi_1\rangle\|$ | $p_{max}^{\psi}$ | $p_{max}^{\rho}$ | $p_{RS}^{\rho}$ | $p_{OAM}$ | $N_{\rho RS}(x)$ | $N_{OAM}(x)$ | Dose(1/Å$^2$) |
|---|---|---|---|---|---|---|---|---|
| Pa,Pb | 0.987 | 0.582 | 0.580 | 0.531 | 0.541 | 346, 1040 | 224, 739 | 0.007,0.023 |
| Pa,Pc | 0.981 | 0.598 | 0.596 | 0.540 | 0.564 | 257, 845 | 98, 323 | 0.003,0.010 |
| Pb,Pc | 0.975 | 0.612 | 0.610 | 0.552 | 0.559 | 143, 468 | 108, 356 | 0.003,0.011 |

**Table 1** Discrimination probabilities based on the use of a single-electron probe and corresponding to different pairs of proteins ($X_0$, $X_1$). The probabilities with superscript $\psi$ and $\rho$ correspond respectively to the case of pure and mixed electron states; the subscripts max, RS, and OAM identify respectively the theoretical maxima, to measurements performed in real space (with an ideal Zernike phase plate) and with the OAM and optimal radial projector . In all the considered cases, the a priori probabilities are assumed equal ( $p_0 = p_1 = 1/2$ ). The values of $N_{\rho RS}$ and $N_{OAM}(x)$ indicates the number of electron necessary to discriminate the proteins with threshold value x=0.9 and 0.99,using real space and optimized OAM sorter, The last column indicated the corresponding doses for $N_{OAM}(x)$.

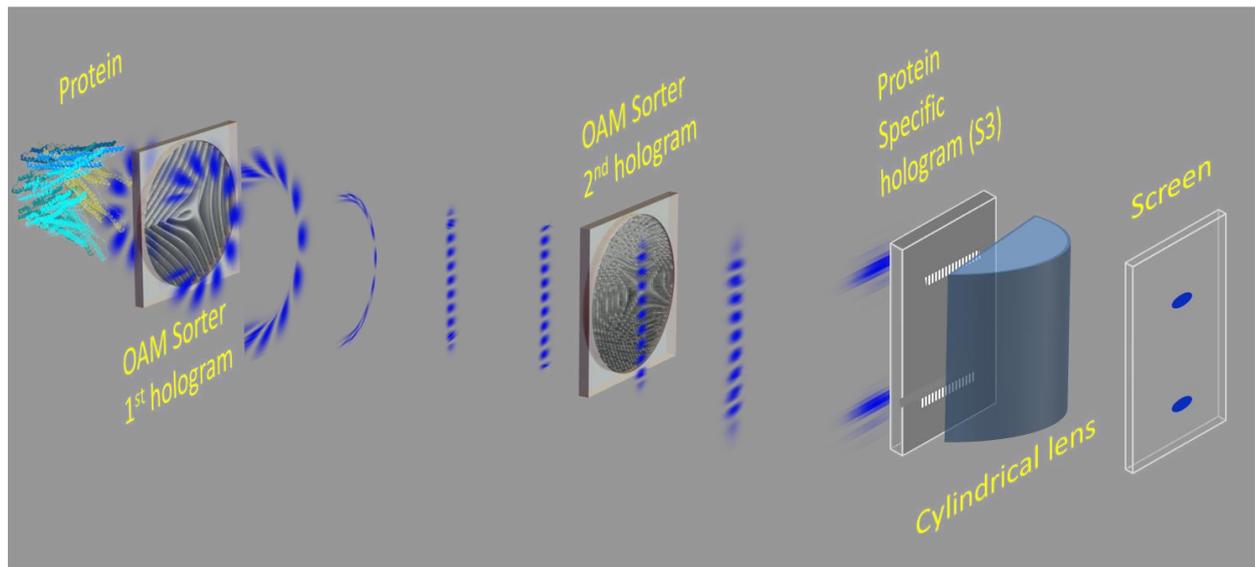

**Fig 1.** Schematic view of the generalized OAM sorter. With respect to the standard OAM sorter, a third phase element is added. The structure of such third element is adapted to the radial structure of the proteins to be recognized.

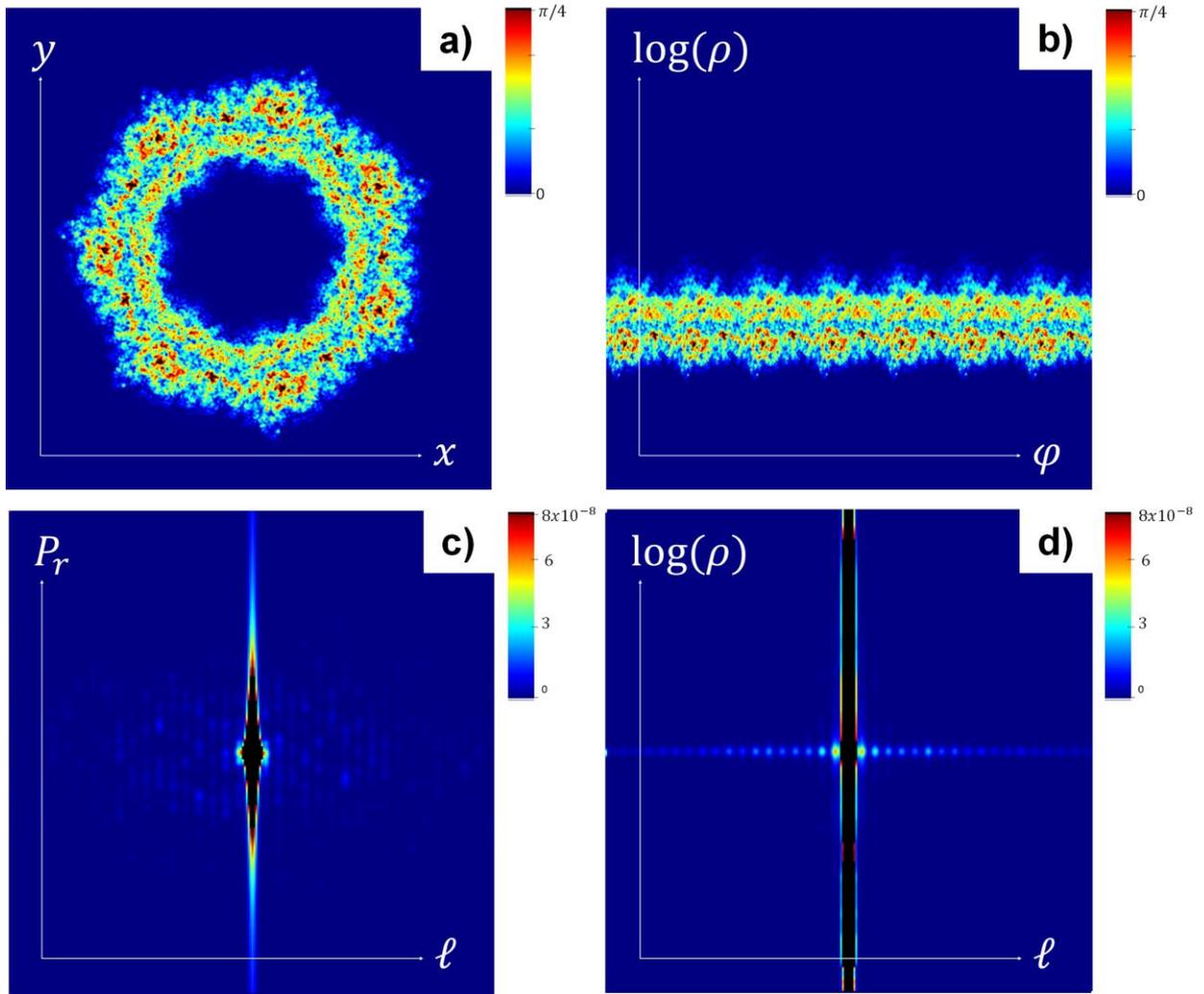

**Fig 2.** Series of unitary transformation that the different electrostatic elements apply to the electron wave function in the case of the protein Pa. In particular, we show: the phase modulation of the electron wave function in (a) Cartesian and (b) log-polar coordinates; (c) the intensity of the diffraction to the OAM sorter in (c) the radial momentum space and (d) after optimal radial sorting.

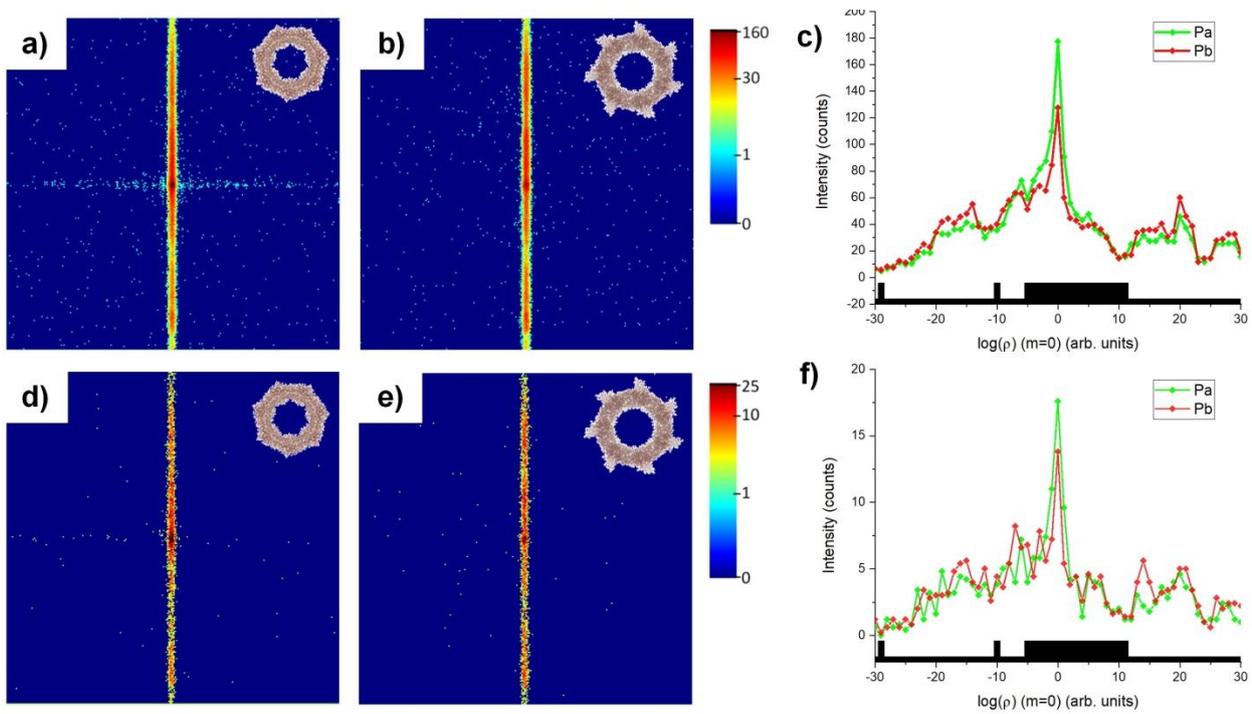

**Fig 3.** Discrimination between the proteins Pa and Pb (both characterized by a 7-fold rotational symmetry), where the third sorter element implements an optimal measurement. The upper and the lower panels correspond respectively to doses of 2 and 0.2 e$^-$/ Å$^2$. The proteins Pa (a,d) and Pb (b,e) give rise to different distributions of the detected electrons in the correlated angular-radial basis. (c,f) Statistics of the radial observable corresponding to the *m=0* subspace, where the electron states are mainly concentrated. The black histogram below the plots represents the optimal projector for Pa: all electron falling on this black pixels are interpreted as indication of Pa.